\documentstyle[sprocl,epsf]{article}

\begin{document}

\title{NA49 RESULTS ON SINGLE PARTICLE AND CORRELATION MEASUREMENTS
IN CENTRAL PB+PB COLLISIONS}
\author{F.~WANG~\footnote{Invited talk given at 
the XXVIII International Symposium on Multiparticle Dynamics,
Delphi, Greece, September 6--11, 1998}}
\address{Lawrence Berkeley National Lab, Berkeley, CA 94720, USA\\
E-mail: FQWang@lbl.gov}
\author{for the NA49 Collaboration:\\
H.~Appelsh\"{a}user$^{7,\#}$, J.~B\"{a}chler$^{5}$,
S.J.~Bailey$^{17}$, D.~Barna$^{4}$, L.S.~Barnby$^{3}$,
J.~Bartke$^{6}$, R.A.~Barton$^{3}$, H.~Bia{\l}\-kowska$^{15}$,
A.~Billmeier$^{10}$, C.O.~Blyth$^{3}$
% ...
}
\address{
% ...
$^{3}$Birmingham University, Birmingham, England.\\
$^{4}$KFKI Research Institute for Particle and Nuclear Physics, Budapest, Hungary.\\
$^{5}$CERN, Geneva, Switzerland.\\
$^{6}$Institute of Nuclear Physics, Cracow, Poland.\\
$^{7}$Gesellschaft f\"{u}r Schwerionenforschung (GSI), Darmstadt, Germany.\\
% ...
$^{10}$Fachbereich Physik der Universit\"{a}t, Frankfurt, Germany.\\
% ...
$^{15}$Institute for Nuclear Studies, Warsaw, Poland.\\
% ...
$^{17}$Nuclear Physics Laboratory, University of Washington, Seattle, WA, USA.\\
% ...
$^{\#}$present address: Physikalisches Institut, Universitaet Heidelberg, Germany.\\
% ...
}
\maketitle

\abstracts{
Single-particle spectra and two-particle correlation functions measured 
by the NA49 collaboration in central Pb+Pb collisions at 158 GeV/nucleon
are presented.
These measurements are used to study the kinetic and chemical freeze-out
conditions in heavy ion collisions. We conclude that large baryon stopping,
high baryon density and strong transverse radial flow are achieved in
central Pb+Pb collisions at the SPS.
}

\section{Introduction}

The goal of studying high energy heavy-ion collisions is to
understand the properties of nuclear matter under extreme conditions
of high energy density ($>1$~GeV/fm$^3$).
It is unlikely under such conditions that spatially separated hadrons
exist. Instead, formation of quark-gluon-plasma (QGP), in which
quarks and gluons are deconfined over an extended volume,
is expected.~\cite{Lee74:qgp,Raf82:qgp,McL86:qgp,qm91:Mul}
This state of matter might have existed in the early universe shortly
after the Big Bang.~\cite{Kei76:star}

If a QGP is produced in a heavy ion collision, the collision system
will undergo transitions from hadrons (initial nucleons in the
incoming nuclei), to partons, to interacting hadrons, to, finally,
freeze-out particles where the measurements are realized.
Many signatures of QGP formation have been proposed.~\cite{qm9697}
These include enhanced production of strangeness,~\cite{Koc86:strange}
charm~\cite{Mat86:charm} and leptons,~\cite{Shu78:lepton}
jet quenching,~\cite{Wan92:quench} collective radial
flow,~\cite{Gyu97:hot-spots} etc.
However, none of the signatures is conclusive, and many 
can be produced by secondary particle
interactions.~\cite{Sor91:rqmd-meson,Mat89:rqmd-k-pi}
Therefore, systematic studies of multiple observables are necessary
in the searching of the QGP formation in heavy ion collisions.

CERN SPS experiment NA49 has measured many observables of Pb+Pb
collisions at 158 GeV/nucleon. These measurements include transverse
energy production,~\cite{Alb95:na49_prl} baryon stopping and negative
hadron distributions,~\cite{App98:na49_prl_baryon} negative
hadron~\cite{App98:epj_expansion} and proton~\cite{App99:na49_2p_correl}
two-particle correlation functions, and charged and
neutral strange particle distributions~\cite{Bor97:sqm97,App98:na49_xi}
in central collisions, and directed and elliptic flow in non-central
collisions.~\cite{App98:na49_prl_flow} In this talk, only selected
results from central Pb+Pb collisions are presented.
Distributions of net baryons, negative hadrons,
and charged kaons are presented in Sect.~\ref{sect:single} to address the
chemical freeze-out conditions. Negative hadron and proton two-particle
correlation functions are presented in Sect.~\ref{sect:two-particle} to
address the kinetic freeze-out conditions.
Conclusions are drawn in Sect.~\ref{sect:conclusion}.
The results which have been published or have been submitted for
publication are not duplicated here but are referred to.
All the plots shown here are preliminary.

NA49 is a large acceptance hadron experiment,~\cite{Wen98:na49_nim} 
measuring about 800 charged particles
in a central Pb+Pb event. The experimental apparatus consists of four
large time-projection chambers (TPCs).
Two of them are placed along the beam axis inside two dipole magnets,
which have maximum integrated field strength of 9 Tesla-meters.
The other two are downstream of the magnets on either side of the beam axis.
Further downstream are four time-of-flight (TOF) walls
covering smaller acceptance than the TPCs.
The particle identification was performed by measuring the specific ionization
(dE/dx) deposited by charged particles in the TPCs,~\cite{Las98:na49_nim}
and the velocities of the particles~\cite{Pue98:qm97_na49_phi} in the
TOF walls. The dE/dx resolution was 5\%, and the TOF resolution was 60 ps.
A beam of Pb nuclei struck on a Pb target of thickness
224 $mg/cm^2$ placed in front of the first TPC.
A zero-degree hadronic calorimeter,~\cite{Alb95:na49_prl,App98:na49_spectator}
located downstream along the beam-axis, measured
mainly the energy remnant in the projectile spectators. By applying a
threshold on the measured energy as an experimental trigger, the 5\%
most central events were recorded. These events corresponded to
collisions with impact parameter below 3.3 fm.
 
\section{Single-particle distributions}
\label{sect:single}

Extensive studies of Si+A and Au+Au collisions at the BNL AGS and
S+A collisions at the CERN SPS have shown that nucleus-nucleus collisions
are not mere superpositions of nucleon-nucleon collisions (N+N) at
the corresponding energies.~\cite{qm9697} Protons are shifted further
into the central rapidity region (baryon stopping) in these
collisions~\cite{Ahl98:proton,Alb98:na35_epj,Bac94:na35_prl,Bac96:sqm96}
than in N+N~\cite{Agu91:pp}.
The NA49 result on the net baryon 
rapidity distribution in central Pb+Pb collisions over a wide rapidity range
(Fig.1 in Ref~\cite{App98:na49_prl_baryon}) shows a larger average
rapidity shift than that in S+S 
(3\% centrality).~\cite{Bac94:na35_prl,Bac96:sqm96}
This is consistent with the picture of more multiple
collisions suffered by each nucleon on average in Pb+Pb than in S+S.

\begin{figure}[hbt]
\centerline{\epsfxsize=3.8in\epsfbox[0 280 600 640]{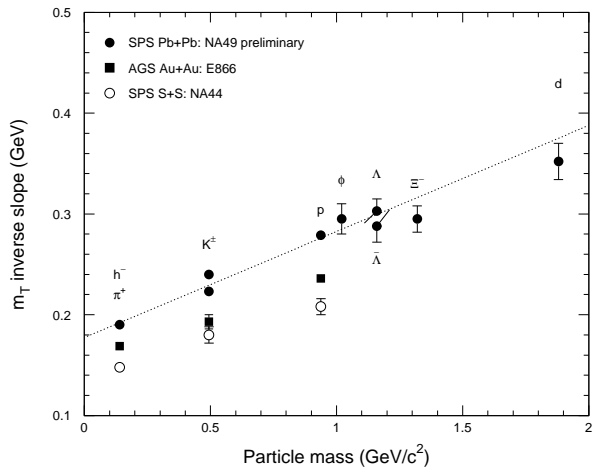}}
\caption{Mass systematics of inverse slopes of particle $m_{\perp}$
spectra in the central rapidity region for three central collision systems:
Pb+Pb (5\%)
at $\sqrt{s}=17.3$~GeV, S+S (3\%)
at $\sqrt{s}=19.4$~GeV, and Au+Au (4\%)
at $\sqrt{s}=4.74$~GeV. The straight line is to guide the eye.}
\label{fig:t_vs_mass}
\end{figure}

A larger fraction of energy is converted from the longitudinal direction
into the transverse
direction in Pb+Pb central collisions than in S+S.
Two consequences can be expected: more particle production and/or
stronger transverse expansion in Pb+Pb than S+S.
Products from high energy heavy ion collisions are dominantly pions.
The rapidity distribution of negative hadrons ($h^-$),
mainly $\pi^-$'s, scales with number of participants 
from S+S to Pb+Pb central collisions
almost perfectly (Fig.2 in Ref~\cite{App98:na49_prl_baryon}),
in contrast to the different net baryon distributions.
Since the total pion multiplicity provides a good approximation to the
total entropy produced, the scaling implies that no extra entropy per
participant was generated in Pb+Pb central collisions with respect to S+S.
However, one should note that pion production does not scale
from N+N to S+S.~\cite{Rol98:qm97}
Collective transverse expansion is reflected in particle transverse
distributions; heavy particles pick up more transverse momentum from
the expansion than light particles, exhibiting larger inverse slopes of
the exponential transverse mass ($m_{\perp}=\sqrt{p^2_{\perp}+m^2}$) spectra.
In Fig.~\ref{fig:t_vs_mass}, the inverse slopes are plotted against 
particle masses ($m$) for three systems: Pb+Pb~\cite{Bor97:sqm97} and
S+S~\cite{Bea97:na44_prl} at the SPS,
and Au+Au at the AGS.~\cite{Ahl98:proton,Wan96:columbia}
The inverse slope systematically increases with particle mass in all
three systems, consistent with the picture of collective
transverse radial flow. The inverse slope is larger in heavy system
Pb+Pb than light system S+S at similar energies, and is larger at high
energy (SPS) than low energy (AGS) in systems with similar sizes.


Kaons, carrying most of the strangeness produced in heavy ion collisions
at SPS energy and lower, have been extensively studied. NA49 has measured
charged kaon production at mid-rapidity by the TOF walls and over wide
rapidity range by the TPCs. The kaon $m_{\perp}$ spectra, 
can be well described by single exponential functions. 
At mid-rapidity, the inverse slopes are measured to be $240\pm 4$~MeV 
for $K^+$ and $223\pm 3$~MeV for $K^-$,
and the yields per unit rapidity are estimated to be $dN/dy=29.8\pm 0.9$
($K^+$) and $16.2\pm 0.5$ ($K^-$).

\begin{figure}[hbt]
\centerline{\epsfxsize=3.8in\epsfbox[0 280 600 640]{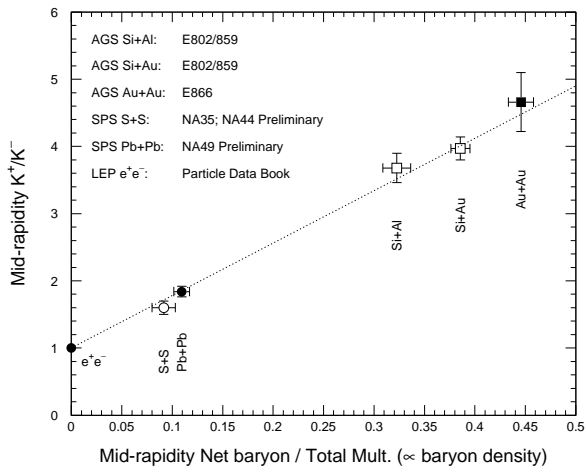}}
\caption{Mid-rapidity $K^+/K^-$ ratio as function of the mid-rapidity ratio
of net-baryon yield over total multiplicity in heavy ion collisions at the
AGS and the SPS. Rapidity integrated $K^+/K^-$ ratio in $e^+e^-$ at LEP 
is also shown. The straight line is to guide the eye.}
\label{fig:kaon_ratio_density}
\end{figure}

The $K^+/K^-$ ratio is sensitive to the baryon 
density. In Fig.~\ref{fig:kaon_ratio_density}, the ratio at mid-rapidity
is plotted against the mid-rapidity ratio of net-baryon yield over 
total-multiplicity for Si+A~\cite{Abb94:e802-hadron,Mor94:MIT,qm95:Col}
and Au+Au~\cite{Ahl98:proton,Ahl98:kaon-auau,Wan96:hipags} central collisions
at the AGS and S+S~\cite{Alb98:na35_epj,Bac94:na35_prl,Bog98:prc_kaon}
and Pb+Pb~\cite{App98:na49_prl_baryon} central collisions at the SPS.
The total multiplicity at mid-rapidity is approximated by the sum of
the net baryon yield and $3\langle\pi\rangle$ at mid-rapidity, 
where $\langle\pi\rangle=(\langle\pi^+\rangle+\langle\pi^-\rangle)/2$ 
for the AGS data and $\langle\pi\rangle=\langle h^-\rangle$ for the SPS data.
Note that both the abscissa and the ordinate are experimentally
measured quantities. However, the motivation to plot this way is that the
ratio of net-baryon yield over total multiplicity is proportional
to the average baryon density at freeze-out, assuming that the volume 
of the system scales with the total multiplicity. 
The $K^+/K^-$ ratio in $e^+e^-$ at LEP~\cite{PDG98:e+e-mult}
is also plotted. As seen from the plot, the $K^+/K^-$
ratio has strong dependence on the baryon density.
Note that the baryon density in heavy ion collisions
is higher at low energies and in large systems.

\begin{figure}[hbt]
\centerline{\epsfxsize=4in\epsfbox[0 300 600 650]{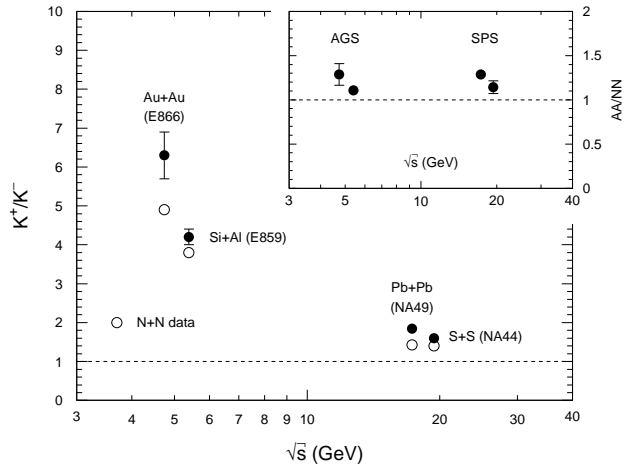}}
\caption{Total yield $K^+/K^-$ ratio (except for Pb+Pb, the
mid-rapidity ratio is used) as function of bombarding energy for N+N
(open points) and heavy ion collisions (filled points).
The ratios of $K^+/K^-$ in heavy ion collisions
over these in N+N are plotted in the insert.}
\label{fig:kaon_ratio}
\end{figure}

In Fig.~\ref{fig:kaon_ratio}, the total yield $K^+/K^-$ ratios in 
heavy ion collisions are compared to those in 
N+N~\cite{Gaz91:nn,Fes79:strangeness,Blo74:pp-mult,Ros75:pp}
at various bombarding energies.
The ratio strongly depends on the bombarding energy in both heavy ion
and N+N collisions. However, the ratios are higher in heavy ion
collisions than in N+N at the same energies. The dependence in N+N can be
understood by the higher production threshold for $K^-$ than $K^+$.
The higher ratio in heavy ion collisions is qualitatively consistent 
with the following conjecture: each nucleon suffers multiple interactions
in heavy ion collisions, where subsequent interactions happen at lower
energy than the primary interaction, producing more $K^+$'s relative
to $K^-$'s. The ratio of $K^+/K^-$ in heavy ion collisions over 
that in N+N (referred as $K^+/K^-$ enhancement) is shown in the 
insert of Fig.~\ref{fig:kaon_ratio}. The enhancement is higher in heavy 
system than in light system, which is consistent with the multiple collision
picture. However, the enhancement seems independent of bombarding energy
in systems with similar sizes.

\section{Two-particle correlation functions}
\label{sect:two-particle}

Two-particle correlation functions provide space-time information 
of the particles at freeze-out,~\cite{Gel90:rmp_corr,Bay98:hbt}
which cannot be obtained from single-particle distributions.
Due to quantum interference, the probability to find two identical pions
(bosons) with low momentum difference is enhanced. The width of the $2\pi$
correlation function after correction for Coulomb interaction is inversely
related to the effective size of the pion source.
Because of the correlation between space-time and momentum of the
particles produced in heavy ion collisions at freeze-out, 
due to the collision dynamics and collective behavior such as flow, 
particles within a limited momentum range are emitted from a limited
space-time region of the source. 
The two-particle correlation function therefore measures, at small momentum
difference, a limited region of the source.
 
NA49 has measured $2h^-$ correlation
functions.~\cite{App98:epj_expansion} 
Based on a model incorporating both longitudinal and transverse
flow,~\cite{Hei96:plb,Cha95:prc,Hei96:qm96_hbt} 
the transverse and longitudinal radii 
were determined to be 6--8~fm at mid-rapidity, dropping by 2--3~fm to 
forward rapidity (Fig.4 in Ref.~\cite{App98:epj_expansion}). 
This size, compared to the 1-dimensional Gaussian
radius $\sigma_{\rm Pb}\approx 3$~fm of the initial nuclei, implies that
the collision system has gone through significant expansion before freeze-out.
The radius parameters decrease with pair transverse momentum $p_{\perp}$
(Fig.6 in Ref.~\cite{App98:epj_expansion}). This behavior was also seen in
S+A collisions~\cite{Alb95:na35_2pi,Bek95:na44_prl}, and is consistent
with transverse radial flow.
Combined with the single-particle transverse distributions, 
the freeze-out temperature (assumed to be the same for all particle species)
and the transverse radial flow velocity at the source surface were
estimated to be $T=120\pm 12$~MeV and $\beta_{\perp}=0.55\pm 0.12$,
respectively.

Similar $p_{\perp}$ dependence of the correlation length has been
found in $Z^0$ hadronic decays in $e^+e^-$ annihilation at
LEP. 
By including all final particles, the LUND model calculation reproduces 
the dependence.~\cite{And98:hbt_lund}
The conjecture is that contribution from resonance decays, dominant at
low $p_{\perp}$, enlarges the correlation length.
The $p_{\perp}$ dependence of the correlation length in $e^+e^-$
may also be due to
space-time-momentum correlation arising from large formation time of
high $p_{\perp}$ particles.
Study of the effects of resonance decays and particle formation time 
on the source size measurement of heavy ion collisions, besides
the effect of transverse radial flow, 
is underway.~\cite{Wan99:2pion_correl_in_pp}


The baryon density plays an important role in the dynamical evolution
of heavy ion collisions.~\cite{Koc83:Strange,Lee88:prc}
To measure baryon density, one needs information on the space-time extent
of the baryon source, which is not necessarily the same as that of the pion's.
The space-time extent of the baryon source at freeze-out
can be measured by the two-proton correlation function. 
The two-proton correlation function has very different characteristics
from the $2\pi$ correlation function.~\cite{Gel90:rmp_corr,Koo77:plb_pp}
Due to Fermi statistics, strong interaction between two protons
is primarily due to 1/4 $S$-wave attractive interaction of the spin
singlet and 3/4 $P$-wave repulsive interaction of the spin triplet.
The combined strong and Coulomb interactions
generate a peak in the two-proton correlation function
at roughly  $q_{\rm inv}=\sqrt{-q_{\mu}q^{\mu}}/2 \approx 20$~MeV/c,
where $q_{\rm inv}$ is the momentum magnitude of either proton in the rest
frame of the pair, and $q_{\mu}$ is the difference of the proton
four-momentum vectors.
The peak height above 1 is inversely proportional to the effective volume
of the proton source.

\begin{figure}[hbt]
\centerline{\epsfxsize=3.8in\epsfbox[0 225 600 580]{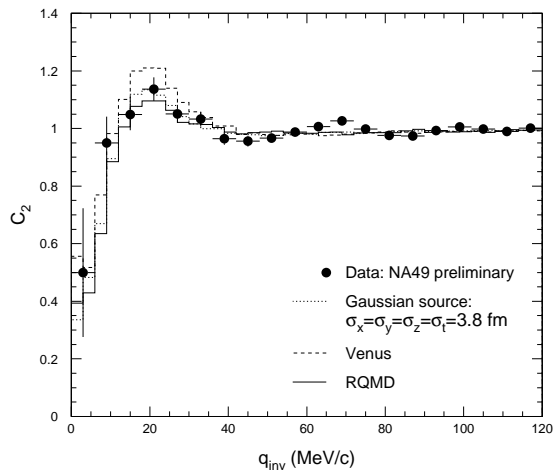}}
\caption{Two-proton correlation functions (not corrected for protons from
weak decays) at mid-rapidity measured by NA49 (points), 
calculated for a Gaussian source (dotted histogram),
and calculated for protons generated by RQMD (solid histogram) 
and by Venus (dashed histogram).}
\label{fig:pp_correl}
\end{figure}

While $2\pi$ correlation function measurements are common, 
measurements of two-proton correlation functions are rare.
NA49 has measured the two-proton correlation function at mid-rapidity.
This is the first such measurement in Pb+Pb collisions at the SPS.
The protons are mostly primordial nucleons, since baryon-antibaryon pair 
production contribution is at the level of 10\% at SPS 
energy.~\cite{Alb96:na35_plb,Bea98:na44_prc_proton}
The measured two-proton correlation function is shown in 
Fig.~\ref{fig:pp_correl}.
The peak at $q_{\rm inv}\approx 20$~MeV/c is evident and the height
is $1.14\pm 0.04$.
The correlation function does not go to zero at $q_{\rm inv}=0$
because of the contamination of protons from weak decays; there,
the effect of the experimental momentum resolution is negligible.
The deviation of the correlation function from 1 at large $q_{\rm inv}$
is currently under investigation.

In order to interpret the proton freeze-out conditions, we employed the
approach of comparing the measured two-proton correlation
function with theoretical calculations. Given the proton phase space
distribution, the two-proton correlation function was calculated using
the Koonin-Pratt Formalism.~\cite{Koo77:plb_pp,Pra87:two_proton}
The experimental momentum resolution was included in the calculation.
For the proton freeze-out distribution, we used two approaches.
(I) Gaussian sources were generated with widths
$\sigma_x=\sigma_y=\sigma_z=\sigma$ for the space coordinates and
$\sigma_t=0$ and $\sigma$ for the time duration. The momentum was 
generated by a thermal distribution of temperature $T=120$~MeV.
The Gaussian sources do not have space-time-momentum correlation.
(II) Protons were generated by two dynamical models,
the Relativistic Quantum Molecular Dynamics (RQMD) model 
(v2.3)~\cite{Sor89:rqmd,Sor92:rqmd-ropes,qm88:Sor:rqmd,Sor95:flavor-changing}
and the Venus model (v4.12).~\cite{Wer93:venus}
The protons have correlation between space-time
and momentum intrinsic to the dynamical evolution of the models.
The strength of the calculated correlation functions for (I) were reduced
by 30\% to account for the contamination of protons from weak decays.
In the calculation for (II),
the experimental acceptance was applied to the protons, and experimental
effects in the accidental reconstruction of weak decay products
as primary protons were incorporated. 

\begin{figure}[hbt]
\centerline{\epsfxsize=3.6in\epsfbox[0 280 600 640]{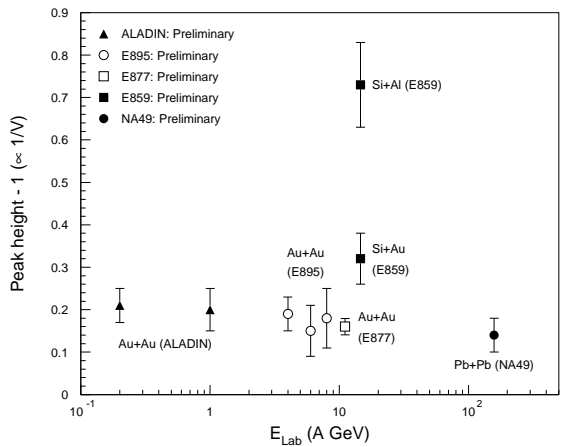}}
\caption{Peak height above 1 of measured two-proton correlation functions
as function of beam energy. Contaminations of protons from weak decays are
not corrected. The correction would bring up the Pb+Pb point by 
approximately 30\%.
The corrections are negligible for data at lower energies.}
\label{fig:pp_correl_peak}
\end{figure}

Comparison between the calculations and the experimental data showed that
the Gaussian source with $\sigma_x=\sigma_y=\sigma_z=\sigma_t=3.8$~fm
well describes the data. The calculated correlation function is plotted
in Fig.~\ref{fig:pp_correl}.
We note that $\sigma=3.8$~fm is slightly larger than the initial 1-dimensional
nuclei size $\sigma_{\rm Pb}\approx 3$~fm, and is significantly smaller than
the radius extracted from the $2h^-$ correlation function.
The 1-dimensional effective size of the Gaussian source, calculated from the
mean separation between protons in the pair rest frame,
is $\sigma_{\Delta}/\sqrt{2}=4.0$~fm.
We have also studied Gaussian sources with extreme
shapes: $\sigma_x=\sigma_y>>\sigma_z$ and $\sigma_x=\sigma_y<<\sigma_z$
(both $\sigma_t=0$). They cannot describe the data satisfactorily.
The calculated two-proton correlation functions for RQMD and Venus are
also shown in Fig.~\ref{fig:pp_correl}.
Venus overpredicts the strength of the correlation function,
while RQMD slightly underpredicts its strength.
The mean separation between protons in the pair rest frame are
($\sigma_{\Delta x},\sigma_{\Delta y},\sigma_{\Delta z})=(6.76,5.34,8.35)$~fm
for RQMD, and $(4.41,4.50,5.44)$~fm for Venus,
where $z$ is the longitudinal coordinate. The 1-dimensional effective size, 
$\sigma_{\Delta}/\sqrt{2}=
\sqrt[3]{\sigma_{\Delta x}\sigma_{\Delta y}\sigma_{\Delta z}}/\sqrt{2}$,
are 4.74~fm and 3.37~fm, respectively for RQMD and Venus.
We note that no simple relation exists between $\sigma_{\Delta}/\sqrt{2}$
and the apparent single-proton source size in the models
due to the dynamics.

We now comment on our two-proton correlation function measurement
in the context of other measurements.
The pion source volume is measured to be proportional to the pion
multiplicity,~\cite{Bea98:na44_prc,Bea97:na44_zpc,Bak96:qm96}
which increases steadily with beam energy in similar colliding
systems.~\cite{Gaz95:pion}
Due to the large pion-nucleon cross-section, one would expect that
protons and pions freeze-out under similar conditions,
therefore, the proton source size would increase with
beam energy as well.
However, our measurement, in conjunction with preliminary results obtained
at GSI~\cite{Sch97:talk,Fri97:talk} and AGS energies,~\cite{e895_prelim}
shows that the peak heights of the correlation functions
in Au+Au and Pb+Pb collisions are rather insensitive to beam energy
(Fig.~\ref{fig:pp_correl_peak}).
This means that the effective volumes of freeze-out protons are similar
in these collisions over the wide energy range.
More detailed studies are needed to understand the instrumental effects
and the change in space-time-momentum correlation as function of 
the beam energy.

\section{Conclusions}
\label{sect:conclusion}

We conclude the following for central Pb+Pb collisions at the SPS:
\begin{itemize}
\itemsep=-1mm
\item	Strong baryon stopping.
	Baryon stopping in Pb+Pb central collisions is stronger than in S+S.
	The baryon spatial density achieved in Pb+Pb central collisions 
	is higher than in S+S at the SPS, 
	and is lower than in Au+Au at the AGS.
	The large baryon density resulted in a large $K^+/K^-$ ratio. 
	The enhancement in the ratio in heavy ion collisions of similar 
	systems over N+N does not depend on the bombarding energy.
\item 	Strong transverse radial flow.
	Inverse slopes of $m_T$ distributions in the central rapidity region
	are larger in Pb+Pb central collisions than in S+S at the SPS and
	in Au+Au at the AGS.
	The radius parameters from $2h^-$ correlation function suggest
	strong expansion.
	The values of freeze-out temperature and transverse radial flow
	velocity are, however, model-dependent.
\item 	Secondary particle rescattering in heavy ion collisions may be 
	the driving force for the large	baryon stopping, the strong 
	transverse radial flow, and the $K^+/K^-$ enhancement.
\item	No excessive entropy production. The negative hadron rapidity
	distribution scales with number of participants from S+S to
	Pb+Pb central collisions.
	No indication of excessive entropy production in Pb+Pb central
	collisions, compared to S+S, is found.
\item	Freeze-out source size. Effective size of the pion source extracted
	from the measured $2h^-$ correlation function is 6--8~fm at 
	mid-rapidity. Effective size of the proton source extracted from
	the measured two-proton correlation function at mid-rapidity is 4~fm.
\end{itemize}


\section*{References}
\bibliography{talk_ref}

\begin{thebibliography}{10}

\bibitem{Lee74:qgp}
T.~D. Lee and G.~C. Wick.
\newblock {\em Phys.~Rev.~D}, 9:2291, 1974.

\bibitem{Raf82:qgp}
J.~Rafelski.
\newblock {\em Phys.~Rep.}, 88:331, 1982.

\bibitem{McL86:qgp}
L.~McLerran.
\newblock {\em Rev.~Mod.~Phys.}, 58:1021, 1986.

\bibitem{qm91:Mul}
B.~M\mbox{\"{u}}ller.
\newblock {\em Nucl.~Phys.}, A544:95c, 1992.

\bibitem{Kei76:star}
B.~D. Keister and L.~S. Kisslinger.
\newblock {\em Phys.~Lett.~B}, 64:117, 1976.

\bibitem{qm9697}
{\em Nucl.~Phys.}, A610, 1996, A638, 1998.
\newblock QM'96 and QM'97 proceedings.

\bibitem{Koc86:strange}
R.~Koch, B.~M\mbox{\"{u}}ller, and J.~Rafelski.
\newblock {\em Phys.~Rep.}, 142:167, 1986.

\bibitem{Mat86:charm}
T.~Matsui and H.~Satz.
\newblock {\em Phys.~Lett.~B}, 178:416, 1986.

\bibitem{Shu78:lepton}
E.~V. Shuryak.
\newblock {\em Phys.~Lett.~B}, 78:150, 1978.

\bibitem{Wan92:quench}
X.-N. Wang and M.~Gyulassy.
\newblock {\em Phys.~Rev.~Lett.}, 68:1480, 1992.

\bibitem{Gyu97:hot-spots}
M.~Gyulassy, D.~H. Rischke, and B.~Zhang.
\newblock {\em Nucl.~Phys.}, A613:397, 1997.

\bibitem{Sor91:rqmd-meson}
H.~Sorge {\it et~al.}
\newblock {\em Phys.~Lett.~B}, 271:37, 1991.

\bibitem{Mat89:rqmd-k-pi}
R.~Mattiello {\it et~al.}
\newblock {\em Phys.~Rev.~Lett.}, 63:1459, 1989.

\bibitem{Alb95:na49_prl}
T.~Alber {\it et~al.}~(NA49~Coll.).
\newblock {\em Phys.~Rev.~Lett.}, 75:3814, 1995.

\bibitem{App98:na49_prl_baryon}
H.~Appelshauser {\it et~al.}~(NA49~Coll.).
\newblock nucl-ex/9810014.

\bibitem{App98:epj_expansion}
H.~Appelshauser {\it et~al.}~(NA49~Coll.).
\newblock {\em Eur.~Phys.~J.}, C2:661, 1998.

\bibitem{App99:na49_2p_correl}
H.~Appelshauser {\it et~al.}~(NA49~Coll.).
\newblock In preparation.

\bibitem{Bor97:sqm97}
C.~Bormann {\it et~al.}~(NA49~Coll.).
\newblock {\em J.~Phys.}, G23:1817, 1997.

\bibitem{App98:na49_xi}
H.~Appelshauser {\it et~al.}~(NA49~Coll.).
\newblock {\em Phys.~Lett.~B}.
\newblock in press, nucl-ex/9810005.

\bibitem{App98:na49_prl_flow}
H.~Appelshauser {\it et~al.}~(NA49~Coll.).
\newblock {\em Phys.~Rev.~Lett.}, 80:4136, 1998.

\bibitem{Wen98:na49_nim}
S.~Wenig {\it et~al.}
\newblock {\em Nucl.~Instr.~and~Meth.}, A409:100, 1998.

\bibitem{Las98:na49_nim}
B.~Lasiuk {\it et~al.}
\newblock {\em Nucl.~Instr.~and~Meth.}, A409:402, 1998.

\bibitem{Pue98:qm97_na49_phi}
F.~Puehlhofer~(NA49 Coll.).
\newblock {\em Nucl.~Phys.}, A638, 1998.

\bibitem{App98:na49_spectator}
H.~Appelshauser {\it et~al.}~(NA49~Coll.).
\newblock {\em Eur.~Phys.~J.}, A2:383, 1998.

\bibitem{Ahl98:proton}
L.~Ahle {\it et~al.}~(E802~Coll.).
\newblock {\em Phys.~Rev.~C}, 57:R466, 1998.

\bibitem{Alb98:na35_epj}
T.~Alber {\it et~al.}~(NA35~Coll.).
\newblock {\em Eur.~Phys.~J.}, C2:643, 1998.

\bibitem{Bac94:na35_prl}
J.~Bachler {\it et~al.}~(NA35~Coll.).
\newblock {\em Phys.~Rev.~Lett.}, 72:1419, 1994.

\bibitem{Bac96:sqm96}
J.~Bachler {\it et~al.}~(NA35~Coll.).
\newblock {\em Heavy Ion Physics}, 4:71, 1996.

\bibitem{Agu91:pp}
M.~Aguilar-Benitez {{\it et~al.}}
\newblock {\em Z.~Phys.~C}, 50:405, 1991.

\bibitem{Rol98:qm97}
G.~Roland~(NA49 Collaboration).
\newblock {\em Nucl.~Phys.}, A638:91, 1998.

\bibitem{Bea97:na44_prl}
I.~G.~Bearden {\it et~al.}~(NA44~Coll.).
\newblock {\em Phys.~Rev.~Lett.}, 78:2080, 1997.

\bibitem{Wan96:columbia}
F.~Wang.
\newblock PhD thesis, Columbia University, 1996.

\bibitem{Abb94:e802-hadron}
T.~Abbott {\it et~al.}~(E802~Coll.).
\newblock {\em Phys.~Rev.~C}, 50:1024, 1994.

\bibitem{Mor94:MIT}
D.~P. Morrison.
\newblock PhD thesis, MIT, May 1994.

\bibitem{qm95:Col}
B.~A.~Cole~(E802 Coll.).
\newblock {\em Nucl.~Phys.}, A590:179c, 1995.

\bibitem{Ahl98:kaon-auau}
L.~Ahle {\it et~al.}~(E802~Coll.).
\newblock {\em Phys.~Rev.~C}.
\newblock In press.

\bibitem{Wan96:hipags}
F.~Wang~(E802 Coll.).
\newblock In {\em Proceedings of Heavy-Ion Physics at the AGS (HIPAGS'96)}.
  Wayne State University, WSU-NP-96-16, 1996.

\bibitem{Bog98:prc_kaon}
H.~B$\o$ggild {\it et~al.}~(NA44~Coll.).
\newblock 1998.
\newblock nucl-ex/9808002.

\bibitem{PDG98:e+e-mult}
Particle~Data Group.
\newblock {\em Eur.~Phys.~J.}, C3:201, 1998.

\bibitem{Gaz91:nn}
M.~Gazdzicki and O.~Hansen.
\newblock {\em Nucl.~Phys.}, A528:754, 1991.

\bibitem{Fes79:strangeness}
H.~Fesefeldt {\it et~al.}
\newblock {\em Nucl.~Phys.}, B147:317, 1979.

\bibitem{Blo74:pp-mult}
V.~Blobel {\it et~al.}
\newblock {\em Nucl.~Phys.}, B69:454, 1974.

\bibitem{Ros75:pp}
E.~Albini {\it et~al.}
\newblock {\em Nucl.~Phys.}, B84:269, 1975.

\bibitem{Gel90:rmp_corr}
D.~H.~Boal {\it et~al.}
\newblock {\em Rev.~Mod.~Phys.}, 62:553, 1990.

\bibitem{Bay98:hbt}
G.~Baym.
\newblock {\em Acta~Phys.~Polon.}, B29:1839, 1998.

\bibitem{Hei96:plb}
U.~Heinz {\it et~al.}
\newblock {\em Phys.~Lett.}, B382:181, 1996.

\bibitem{Cha95:prc}
S.~Chapman, J.~R. Nix, and U.~Heinz.
\newblock {\em Phys.~Rev.~C}, 52:2694, 1995.

\bibitem{Hei96:qm96_hbt}
U.~Heinz.
\newblock {\em Nucl.~Phys.}, A610:264c, 1996.

\bibitem{Alb95:na35_2pi}
T.~Alber {\it et~al.}~(NA35~Coll.).
\newblock {\em Z.~Phys.~C}, 66:77, 1995.

\bibitem{Bek95:na44_prl}
H.~Beker {\it et~al.}~(NA44~Coll.).
\newblock {\em Phys.~Rev.~Lett.}, 74:3340, 1995.

\bibitem{And98:hbt_lund}
B.~Andersson and M.~Ringner.
\newblock {\em Nucl.~Phys.}, B513:627--644, 1998.

\bibitem{Wan99:2pion_correl_in_pp}
F.~Wang, S.~Esumi, and N.~Xu.
\newblock In preparation.

\bibitem{Koc83:Strange}
P.~Koch, J.~Rafelski, and W.~Greiner.
\newblock {\em Phys.~Lett.~B}, 123:151, 1983.

\bibitem{Lee88:prc}
K.~S.~Lee {\it et~al.}
\newblock {\em Phys.~Rev.~C}, 37:1452, 1988.

\bibitem{Koo77:plb_pp}
S.~E. Koonin.
\newblock {\em Phys.~Lett.}, 70B:43, 1977.

\bibitem{Alb96:na35_plb}
T.~Alber {\it et~al.}~(NA35~Coll.).
\newblock {\em Phys.~Lett.~B}, 366:56, 1996.

\bibitem{Bea98:na44_prc_proton}
I.~G.~Bearden {\it et~al.}~(NA44~Coll.).
\newblock {\em Phys.~Rev.~C}, 57:837, 1998.

\bibitem{Pra87:two_proton}
S.~Pratt and M.~B. Tsang.
\newblock {\em Phys.~Rev.~C}, 36:2390, 1987.

\bibitem{Sor89:rqmd}
H.~Sorge, H.~Stocker, and W.~Greiner.
\newblock {\em {\em Annals~of~Physics}}, 192:266, 1989.

\bibitem{Sor92:rqmd-ropes}
H.~Sorge {\it et~al.}
\newblock {\em Phys.~Lett.~B}, 289:6, 1992.

\bibitem{qm88:Sor:rqmd}
H.~Sorge.
\newblock {\em Nucl.~Phys.}, A498:567c, 1989.

\bibitem{Sor95:flavor-changing}
H.~Sorge.
\newblock {\em Phys.~Rev.~C}, 52:3291, 1995.

\bibitem{Wer93:venus}
K.~Werner.
\newblock {\em Phys.~Rep.}, 232:87, 1993.

\bibitem{Bea98:na44_prc}
I.~G.~Bearden {\it et~al.}~(NA44~Coll.).
\newblock {\em Phys.~Rev.~C}, 58:1656, 1998.

\bibitem{Bea97:na44_zpc}
I.~G.~Bearden {\it et~al.}~(NA44~Coll.).
\newblock {\em Z.~Phys.~C}, 75:619, 1997.

\bibitem{Bak96:qm96}
M.~D.~Baker~(E802 Coll.).
\newblock {\em Nucl.~Phys.}, A610:213c, 1996.

\bibitem{Gaz95:pion}
M.~Gazdzicki and D.~Roehrich.
\newblock {\em Z.~Phys.~C}, 65:215, 1995.

\bibitem{Sch97:talk}
C.~Schwarz~{{\it et~al.}}~(ALADIN Coll.).
\newblock nucl-ex/9704001.

\bibitem{Fri97:talk}
R.~Fritz~{{\it et~al.}}~(ALADIN Coll.).
\newblock nucl-ex/9704002.

\bibitem{e895_prelim}
S.~Panitkin {\it et~al.}~(E895~Coll.).
\newblock E895 preliminary.

\end{thebibliography}

\end{document}